\documentclass[twoside,english,review,3p]{elsarticle}
\usepackage{array}
\usepackage{multirow}
\usepackage{amsmath}

\makeatletter

\providecommand{\tabularnewline}{\\}

\usepackage{amsmath}
\usepackage{accents}

\usepackage{amsfonts}
\usepackage{mathtools}
\usepackage{mathalfa}

\usepackage{mathrsfs, stmaryrd}
\usepackage{amssymb}
\usepackage{listings}
\usepackage{mathrsfs}
\usepackage{times}
\usepackage{float}
\usepackage{setspace}
\usepackage{color,soul}

\usepackage{amsmath,amsfonts,amsthm,eucal}
\usepackage{enumerate}
\usepackage{float}
\usepackage[title]{appendix}
\usepackage{color}
\usepackage{caption}
\usepackage{subcaption}
\usepackage{graphicx}

\usepackage{booktabs} 
\usepackage{lineno}

\usepackage{adjustbox}

\usepackage{algorithm}
\usepackage{bm}
\usepackage{algpseudocode}%[noend]

\def\beq{\begin{equation}}
\def\eeq{\end{equation}}
\def\bal{\begin{equation} \begin{aligned}}
\def\eal{\end{aligned} \end{equation}}

\providecommand{\blue}[1]{#1}
\providecommand{\bs}[1]{\ensuremath{\boldsymbol{#1}}}

\providecommand{\dd}[2]{\frac{\partial #1}{\partial #2}}

\providecommand{\dev}[1]{\,\text{dev}\left[#1\right]}
\providecommand{\vol}[1]{\,\text{vol}\left[#1\right]}
\renewcommand{\exp}[1]{\,\text{exp}\left[#1\right]}

\DeclareMathOperator{\dl}{\Delta \lambda}

\theoremstyle{remark}

\DeclarePairedDelimiter\norm{\lVert}{\rVert}%

\makeatletter

\newcommand{\volumedash}{%
  \makebox[0pt][l]{%
    \ooalign{\hfil\hphantom{$\m@th V$}\hfil\cr\kern0.08em--\hfil\cr}%
  }%
}
\makeatother

\journal{LA-UR-23-34028}
\makeatother

\usepackage{babel}

\begin{document}

\begin{frontmatter}{}

\title{An integrated EOS, pore-crush, strength and damage model framework for near-field ground-shock}

\author[essd]{K.C.~Bennett}
\ead{kaneb@lanl.gov}

\author[essd]{A.M.~Stahl}
\author[td]{T.R.~Canfield}
\author[essd]{G.G.~Euler}

\cortext[cor1]{Corresponding author. Earth and Environmental Sciences Division,
Los Alamos National Laboratory, Los Alamos, New Mexico. }

\address[essd]{Earth and Environmental Sciences Division, Los Alamos National
Laboratory, Los Alamos, New Mexico}

\address[td]{Theoretical Division, Los Alamos National
Laboratory, Los Alamos, New Mexico}

%ABSTRACT
\begin{singlespace}
\begin{abstract}
An integrated Equation of State (EOS) and strength/pore-crush/damage model framework is provided for modeling near to source (near-field) ground-shock response, where large deformations and pressures necessitate coupling EOS with pressure-dependent plastic yield and damage. Nonlinear pressure-dependence of strength up to high-pressures is combined with a Modified Cam-Clay-like cap-plasticity model in a way to allow degradation of strength from pore-crush damage, what we call the ``Yp-Cap'' model. Nonlinear hardening under compaction allows modeling the crush-out of pores in combination with a fully saturated EOS, i.e., for modeling partially saturated ground-shock response, where air-filled voids crush. Attention is given to algorithmic clarity and efficiency of the provided model, and the model is employed in example numerical simulations, including finite element simulations of underground explosions to exemplify its robustness and utility.
\end{abstract}
\end{singlespace}

%KEYWORDS

\end{frontmatter}{}

%%. **********.  END FRONTMATTER *************

% SECTION 1
\section{Introduction\label{sec:introduction}}
Impacts and blasts in and on the Earth induce large deformations and stresses that dissipate with distance from the source \citep{johnson2001effects, steedman2016phenomenology, mellors2018source, vorobiev2023geodyn}. Equations of state (EOS) are often required to accurately model the nonlinear pressure, density and temperature relations of geomaterials near to the source (near-field) \citep{trunin1994shock}, but strength (plastic yielding) and damage mechanisms of deformation are also often evident in the near-field \citep{grady2000computational, froment_etal_2020, stahl2022probing}---necessitating the incorporation of EOS within geomechanical models of elastoplasticity and damage. 

Significant advancements in geomaterial EOS were made in the later half of the last century, especially associated with modeling of underground nuclear tests \citep[e.g.,][]{furnish1993recent, erskine1994shock}. The importance and challenge of including the effect of air-filled porosity crushing out, i.e., ``pore-crush,'' has also been identified for some time, with various approaches proposed \citep{Margolin1986, carroll1972static, menikoff2000equation}. Improvements in geomaterial EOS continues to be a topic of active research, especially as measurement techniques advance \citep{crockett2021equation}. Notably, better descriptions of temperature \citep{rubin2020analytical, strzelecki2022determination}, multiphase mixtures and phase changes \citep{schoelmerich2020evidence, gammel2022CHEST}, and the influence of dry and water-filled porosity \citep{lajeunesse2017dynamic, vogler2019shock} remain active areas of research. 

Geomechanics models for elastoplasticty and damage have of course also been a topic of much development. Pressure dependency of strength has long been well-understood \citep{drucker1952soil,duncan1970nonlinear}, but continues to be developed, e.g., for multiscale modeling \citep{bryant2022multiscale}, coupling to damage \citep{brannon2009survey, BennettBorja18, xie_etal_2017}, anisotropy \citep{vorobiev2008generic, semnani2016thermoplasticity, bennett2019anisotropic} and for various other needs \citep{lee2023multilevel}. Similarly, computational models inclusive of compactive yielding have been well elucidated \citep[cf.][]{borja1990cam}, but there remain various topics of active development, such as coupling to damage \citep{bennett2019energy, hagengruber2022strength, ip2023modeling}, thermodynamic-consistency at large deformation \citep{Bennettetal16}, and dynamic applications \citep{vorobiev2007simulation}. Despite these many advancements in geomaterial EOS and geomechanical models, there remains an outstanding need for their integration. Recent work in this regard includes that of
\citep{grady2000computational, banerjeetheory, Cui_etal_2017, vorobiev2023geodyn, TONGE201676}, among others.

Herein, a novel model, what we call the ``Yp-Cap'' model, is presented for integrating EOS for saturated porous rock with nonlinear pressure dependent yield, pore-crushing, and pore-crush induced damage. The model is motivated by the desire to utilize the Sesame tabular EOS database of Los Alamos National Laboratory \citep{sesame} in conjunction with geomaterial models for near-field response of the Earth to surface and underground explosions. The nonlinear pressure dependent strength model proposed by \citet{kamm1995comparison}, the ``Yp'' (pronounced as an initialism) model is combined with a Modified Cam-Clay (MCC) \citep[cf.,][]{borja1990cam, bennett2019energy} like yield surface in order to include the effect of pore-crush as plastic compression. A novel expression for a MCC-like yield surface is posited that allows hardening of the pressure-cap in compaction without hardening in dilation. The merging of the two surfaces is done in such a way that damage in deviatoric strength from pore-crush can be induced while the pressure-cap is hardening, a phenomena observed to occur under high-pressure shock loading of rocks and concrete \citep[e.g.,][]{Cui_etal_2017}. The partition of volumetric strains into elastic and plastic parts within the return mapping equations is done in a manner thermodynamically consistent with the Sesame EOS database, but without need to perform (computationally costly) interpolations of the Sesame tables within the iterative process. Analysis of the model results for pure-pressure loading is provided to show the efficacy in coupling saturated EOS with pore-crushing to account for partial saturation where only the air-filled portion of the porosity crushes out. The robustness and utility of the material model is demonstrated in simulation of near-field ground-shock response of partially saturated tuff.

%SECTION 2
\section{Assumptions and constitutive setting\label{sec:methods}}
The model is designed particularly for use with the Sesame tabular EOS database, and though we limit the presentation of the algorithms to use of the Sesame tables for the sake of simplicity, we note that the model is easily adaptable to any general type of EOS in either tabular or analytical (e.g., Mie--Gr{\"u}neisen) form. An important distinction of the model, however, is that it assumes the EOS does not incorporate the effect of pore-crush, i.e., the EOS is for fully fluid-saturated porosity. This is because the air-filled porosity is alternatively handled through the pore-crush modeling, as will be subsequently described. This approach is advantageous in that the EOS need not cover a range of partial saturations \citep{butkovich1973technique}---natural variability of geomaterials is challenging enough to encompass---and that the pore-crush can be deconvoluted from the EOS for analysis.

We formulate the model firstly in terms of small strain, which makes for clear description of its novelty (our main objective here). It is important to state, however, that further (standard) treatment is necessary for applications to large strain, which we provide a brief overview of in Section \ref{sec:large-def}. Namely, identification of the Kirchhoff stress and logarithmic strain for hyperelastic formulation, or an incremental constitutive relation and some care ensuring objectivity for a hypoelastic one are required. We also limit the model presentation and example results to rate-independence, which, although it simplifies the description, we note is a limitation for covering a range of rates with a single set of material parameters. Ground shock response is observed to be generally rate-dependent \citep{furnish1993planar, grady1998shock}, and integrating of the rate-independent Sesame EOS tables with rate-dependent pore-crush and strength is part of our ongoing research. 

We start by identifying the per-unit volume free energy of the material (also called ``Helmholtz free energy"), $\Psi$, as a function of the elastic strain, $\bs{\varepsilon}^e$, temperature, $T$, and strain-like internal state variable (ISV), $z$, i.e.,
\bal \label{eq:Psi}
\Psi &= \hat{\Psi}(\bs{\varepsilon}^e, T, z), \\
\Psi &= \hat{\Psi}^v(\vol{\bs{\varepsilon}^e}, T) + \hat{\Psi}'(\dev{\bs{\varepsilon}^e}, T) + \hat{\Psi}^s(z, T),
\eal
where we assume additive decomposition of volumetric, $\Psi^v$, deviatoric, $\Psi'$, and stored energy, $\Psi^s$. The stress is thus defined with respect to the elastic strain energy, $\bs{\sigma}(\bs{\varepsilon}^e, E) := \partial \Psi/\partial \bs{\varepsilon}^e$, where $E$ is the specific internal energy, the functional dependence of which comes through standard thermodynamic relations between entropy, temperature, $\Psi$, and $E$ \citep[cf.,][]{truesdell2004non, menikoff2000equation}. Similarly, the stress-like ISV $\beta$ is defined with respect to the derivative of the stored energy potential, i.e., $\beta := \partial \Psi/\partial z$. The precise form of $\Psi$ will be given in Section \ref{sec:theory}. 

The yield function, $\mathcal{F}$, is assumed generally a function of the Cauchy stress, $\bs{\sigma}$, the stress-like ISV $\beta$, and temperature $T$,
\beq
\mathcal{F} = \hat{\mathcal{F}}(\bs{\sigma}, \beta, T).
\eeq
It describes a surface (``yield surface'') that bounds the elastic domain, $\mathbb{E}$, i.e., the admissible states of $\{\bs{\sigma}, \beta  \}$ are bounded by $\partial\mathbb{E}_\sigma:= \{(\bs{\sigma},\beta) \in \mathbb{S} \times \mathbb{R}\, | \, \mathcal{F}\left( \bs{\sigma}(\bs{\varepsilon}^e), \beta(z, T) \right) = 0\}$. 

The plastic dissipation inequality describes a constraint on the dissipation of energy, $\mathcal{D}$, that arises from the second law of thermodynamics \citep[cf.,][]{Bennettetal16,bennett2019energy},
\beq \label{eq:dissipation}
\mathcal{D} := \dot{\bs{\varepsilon}}^p{:} \bs{\sigma} - \dot{z}\beta \geq 0,
\eeq 
where the strain rate, $\dot{\bs{\varepsilon}}=\dot{\bs{\varepsilon}}^e + \dot{\bs{\varepsilon}}^p$, is additively decomposed into elastic, $\dot{\bs{\varepsilon}}^e$, and plastic $\dot{\bs{\varepsilon}}^p$, parts. The stress and strain are decomposable into volumetric and deviatoric parts in the standard way, e.g., $\bs{\sigma} = \dev{\bs{\sigma}} + \vol{\bs{\sigma}} = \bs{s} +  p\bs{1}$, where the second order identity $\bs{1}$ is identified, along with the mean stress $p$ and the deviatoric stress $\bs{s}$. We use standard mechanics convention for tension positive, but with pressure, $P$, opposite sign as mean stress, i.e., $P=-p$. We use the notation $\bs{\varepsilon} = \bs{\varepsilon}' + e_v/3\, \bs{1}$ for the volumetric/deviatoric decomposition of the strain, with $e_v$ the volume strain. \blue{As will be discussed in the next section, plastic volumetric strain, $e^p_v$, occurs in the model presently only in compaction.}

The flow rule and ISV evolution equations follow from Eq.~\eqref{eq:dissipation} as
\begin{subequations} \label{eq:evols}
\beq
\dot{\bs{\varepsilon}}^p = \dot{\lambda} \dd{\mathcal{F}(\bs{\sigma}, \beta)}{\bs{\sigma}} \quad \implies \quad \Delta{\bs{\varepsilon}}^p = \Delta{\lambda} \dd{\mathcal{F},(\bs{\sigma}, \beta)}{\bs{\sigma}}
\eeq
\beq
\dot{z} = -\dot{\lambda} \dd{\mathcal{F}(\bs{\sigma}, \beta)}{\beta} \quad \implies \quad \Delta z = -\Delta{\lambda} \dd{\mathcal{F}(\bs{\sigma}, \beta)}{\beta},
\eeq
\end{subequations}
where $\dot{\lambda}$ is the ``plastic multiplier,'' and where the discrete rate notation is used, e.g., $\dot{z} = \Delta z/\Delta t$, with time discretized over time step $(n)$ to $(n+1)$ being $\Delta t = t_{(n+1)} - t_{(n)}$.

%SECTION 3
\section{Theory\label{sec:theory}}

\subsection{Yield surface\label{subsec:yield}}
\begin{figure}[htbp]
\centering
\small
\begin{subfigure}[b]{0.45\textwidth}
\includegraphics[width=0.9\textwidth]{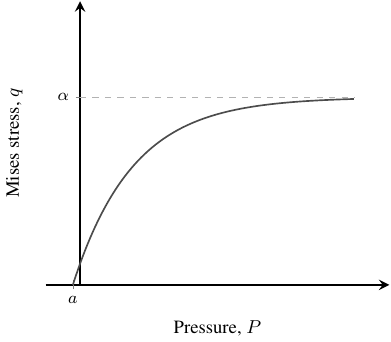}%main-figure0.pdf}
\end{subfigure}
\begin{subfigure}[b]{0.45\textwidth}
\includegraphics[width=0.9\textwidth]{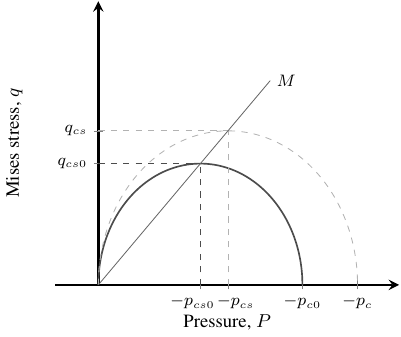}%main-figure1.pdf}
\end{subfigure}
\caption{The two yield surfaces that combine into the Yp-Cap model: (a) The Yp yield surface, showing nonlinear dependence on pressure approaching maximum strength, $\alpha$, at high pressure (strength ``cut-off") and tensile apex, $a$; and (b) the classical MCC yield surface and surface-evolution, showing the CSL with slope $M$, preconsolidation stress $p_{c}$ and critical state points $\{p_{cs},q_{cs}$\}.}
\label{fig:surfs}
\end{figure}

\begin{figure}[htbp]
\centering
\begin{subfigure}[b]{0.45\textwidth}
\includegraphics[width=0.9\textwidth]{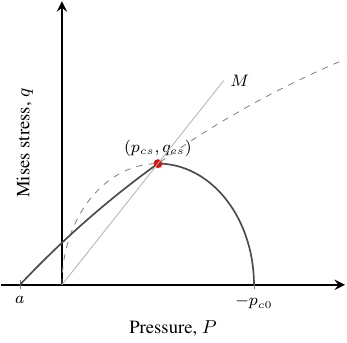}
\end{subfigure} %end subfigure
\hfill
\begin{subfigure}[b]{0.45\textwidth}
\includegraphics[width=0.9\textwidth]{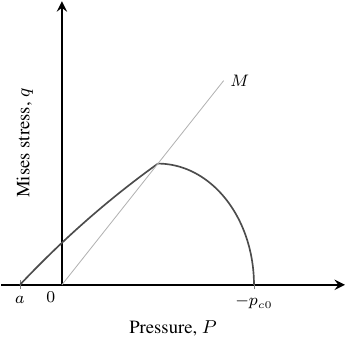}
\end{subfigure} %end subfigure
\caption{Composite Yp-Cap yield surface that is made up of the Yp and MCC surfaces: (a) The intersection of the surfaces at the critical state, and (b) the complete Yp-Cap yield surface.\label{fig:surf-combine}}
\end{figure}

The Yp-Cap yield surface is a single surface composed of two other yield surfaces (Fig.~\ref{fig:surfs}): the Yp nonlinearly pressure dependent surface \citep{kamm1995comparison}, and a novel Modified Cam-Clay (MCC) \citep[cf.,][]{borja1990cam} like surface that is designed to have an evolvable Critical State Line (CSL). The CSL intercept (a.k.a.~the ``brittle-ductile transition'' for rock yield surfaces) with each surface forms the transition from one surface to the other (Fig. \ref{fig:surf-combine}).

The Yp surface (Fig.~\ref{fig:surfs}a) is the same exponential form that was used by \citet{kamm1995comparison}, similar variants of which have been used previously by others~\citep{schwer1994three}. It is designed to cover observed nonlinearity of pressure dependence over large pressure ranges and have a maximum (or ``cut-off'') strength,
\beq \label{eq:Ypfunc}
\mathcal{F}_{\textrm{Yp}} := q - \alpha + \gamma \exp{\frac{p}{P_y}} = 0,
\eeq 
 where $q:= \sqrt{3/2}\norm{\bs{s}}$ is the Mises stress and $p$ the mean stress, and $\alpha\geq 0$, $\gamma\geq 0$, and $P_y \geq 0$ are parameters. The cut-off strength at high-pressure is prescribed by $\alpha$, and the tension cut-off ($a$) can be calculated from Eq.~\eqref{eq:Ypfunc} as $p|_{q=0} = P_y\ln{[\alpha/\gamma]}$. The ratio $R:= \gamma/\alpha$ is restricted $0 < R \leq 1$ to keep the tensile cut-off apex $a$ on the tensile side, where $R=1$ sets $a$ to be at zero pressure ($P=0$). 
 
\begin{figure}[htbp]
\centering
\begin{subfigure}[b]{0.45\textwidth}
\includegraphics[width=0.9\textwidth]{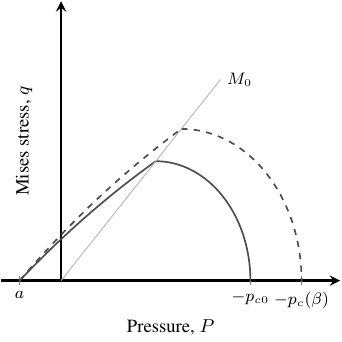}
\caption{$X=0$}
\end{subfigure} %end subfigure
\hfill
\begin{subfigure}[b]{0.45\textwidth}
\includegraphics[width=0.9\textwidth]{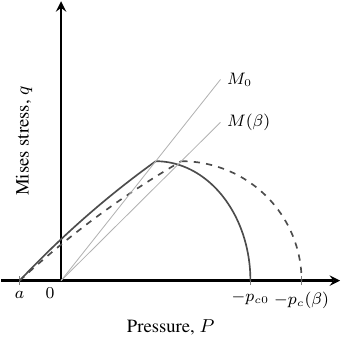}
\caption{$X=1$}
\end{subfigure} %end subfigure
\caption{Evolution of Yp-Cap yield surface for different values of the \emph{propensity for damage parameter}, $X$: (a) $X=0$ and (b) $X=1$.}
\label{fig:evolsurf}
\end{figure}

 The MCC-like surface (Fig.~\ref{fig:surfs}b) is derived such that it can be limited to grow (harden) only on the compactive side of critical state during pore-crush. This is achieved by making the slope of the CSL line a function of the preconsolidation stress, $M = \hat{M}(p_c)$, which will be shown to be a function of the stress-like ISV, $p_c=\hat{p}_{c}(\beta) \implies M=\hat{M}(\beta)$. Flexibility to recover the standard MCC surface is achieved by including the \emph{propensity for damage parameter} $X$,
 \begin{subequations} \label{eq:MCC}
 \beq
 \mathcal{F}_{\textrm{MCC}} := \left(\frac{q}{ M  } \right)^2 + p(p -p_c) = 0
 \eeq
 \beq \label{eq:M}
 M :=   \big(Xp_{c0}/p_c + (1-X)\big)M_0,
 \eeq
 \end{subequations}
where $M_0$ is the initial slope of the CSL and $p_c$ and $p_{c0}$ are the preconsolidation stress and its initial value, for $0\leq X\leq 1$. If $X=0$, there is no degradation of the Yp strength curve (in fact it hardens); if $X=1$, then degradation of strength ensues from pore-crush, limiting the peak critical state strength (Fig.~\ref{fig:evolsurf}), and intermediate values of $X$ interpolate between these extremes.

To merge the surfaces, we require them to intersect on the CSL. To do so, we identify the critical state in $q$-$p$ space (Fig.~\ref{fig:surf-combine}), $\{q_{cs}, p_{cs}\}$, where $q_{cs} = -Mp_{cs} = -Mp_c/2$. We make use of the ratio of parameters of the Yp surface, $R:= \gamma_0/\alpha_0$, and allow $\alpha$ to be a function of $p_c$ as well,
\beq \label{eq:alpha}
\alpha = \frac{ M p_{cs}    }  { R\exp{\frac{p_{cs}}{P_y}} -1   }.
\eeq
The ratio $R$ is required to remain fixed during the simulation, such that during evolution of the surface with ISV $z$, $p_c$ is updated by $\beta=\hat{\beta}(z)$, $M$ by Eq.~\eqref{eq:M}, $\alpha$ from Eq.~\eqref{eq:alpha}, and $\gamma$ by $\gamma = R\alpha$. \blue{This facilitates the co-evolution of the surfaces, but in effect requires $a$ to be fixed, which may be unrealistic for damaging rock. This is mitigated somewhat, however, by degradation of $M$, and it is our view that a load path through $a$ (hydrostatic tension) should be handled by discrete fracture, which is beyond the scope of the current work.} Initial value $M_0$ is calculated from $\{\alpha_0,\gamma_0, p_{c0}\}$ by solving Eq.~\eqref{eq:alpha} for $M$.

%FIGURE 1 Yp and MCC Surfaces

%FIGURE COMPOSITE SURFACE

%FIGURE SURFACE EVOLUTION

\subsection{Elastic strain energy and EOS\label{subsec:yield}}
The Helmholtz free energy Eq.~\eqref{eq:Psi} is identified in its parts. The volumetric part, $\Psi_v$, is calculated by the Sesame tabular EOS as a function of specific volume and temperature. The deviatoric part is quadratic in strain by shear modulus, $G$, i.e., $\Psi' = G \bs{\varepsilon}'{:}\bs{\varepsilon}'$ ($\bs{s}$ is linear in strain by $G$), and the stored energy describes the pore-crush potential through the phenomenological form
\beq \label{eq:storedenergy}
\Psi^s := \frac{Hz^2}{2} + \beta_{\textrm{max}} \left( \frac{\exp{\omega z}}{\omega} - z \right),
\eeq
where $H$ is the linear hardening modulus when exponential parameter $\beta_{\textrm{max}}= 0$, with $\omega$ and $\beta_{\textrm{max}}$ parameters that describe the exponential hardening.

The pressure is thus obtained from the Helmholtz free-energy as a function of specific volume (or density) and temperature through the EOS. The stress-like ISV associated with pore-crush is
\beq \label{eq:beta}
\beta := \dd{\Psi^s}{z} = Hz +\beta_{\textrm{max}} (\exp{\omega z} -1),
\eeq
and we identify the preconsolidation pressure $p_c = p_{c0} -\beta$, where $\beta \geq 0$, $z \geq 0$, and $\Psi^s$ is a positive convex function of $z$.

\subsubsection{Saturated EOS with pore-crush\label{sec:satEOS}}
The EOS, whether Sesame or any other, is required to be representative of the saturated geomaterial (if it is porous). The application to partially saturated geomaterials thus requires some explanation. From basic mixture theory, a representative volume element (RVE) of material is composed of solid, liquid and gas volume fractions ($\phi^s, \phi^l, \phi^g$, respectively). We assume the porosity, $\phi^{l+g}$, is composed of the liquid and gas phases,
\beq
\phi^{l+g} = \phi^l + \phi^g \quad , \qquad \phi^s + \phi^l +\phi^g = 1.
\eeq
We identify the density of a representative volume ($v$) element (RVE) of mass ($m$) composed of a mixture of solid, liquid and gas in the standard way \citep{borja2006mechanical}, 
\beq
\rho = \frac{m}{v} = \frac{m^s + m^l + m^g}{v^s+v^l + v^g} = \frac{m^s + m^l }{v^s+v^l + v^g},
\eeq 
where the mass of the gas is assumed negligible ($m^g \approx 0$). This is the actual (total) density of the material. We then identify the density associated with just the solid and liquid phases as
\beq
\rho^{s+l} = \frac{m^{s+l}}{v^{s+l}} = \frac{m^s + m^l }{v^s+v^l },
\eeq 
which differs from the total density by the volume of the gas phase and is what is used in the EOS calculations. The evolution of density over an increment of volume strain \citep[cf.,][]{bennett2019energy} is thus
\beq
\rho_{(n+1)} = \frac{\rho_{(n)}}{1 + \Delta e_v} \quad , \quad \rho^{s+l}_{(n+1)} = \frac{\rho^{s+l}_{(n)}}{1 + \Delta e^e_v} \, \,,
\eeq
where the partition of elastic and plastic volume strains are determined according to the plastic yield criteria and return mapping equations subsequently described. \blue{Although it could be possible to model compression of the air filled voids with an EOS for air and develop a mixed gas-liquid phase effective stress law, we do not do so here for two reasons based upon our point of view. The first is that there is a lack of experimentally-based understanding of exactly how to formulate such an effective stress law for dynamic states, and since the contribution of pressure from the gas phase is relatively much less than the liquid (because it is so much more compliant), the cost of tracking it isn't justifiable \citep[cf.][]{irwin2024large}. The second is that, although the effect of local temperature fluctuations from compressing air voids may be significant, this phenomena is also not well characterized nor understood in geomaterials \citep[cf.][]{lajeunesse2017dynamic}.} The initial total density of the composite rock is tracked for the global boundary value problem solution. It follows from the previous equations that total density is related to the solid-liquid phase density and the gas-filled porosity by
\beq
\rho_0 = \rho_0^{s+l} (1-\phi^g).
\eeq

%SECTION 4
\section{Numerical implementation and algorithms\label{sec:implementation}}
The Yp-Cap material model is implemented for demonstration in the commercial Finite Element Method (FEM) code Abaqus Explicit \citep{Abaqus2011} utilizing Abaqus's ``VUMAT'' material model subroutine format \citet{Abaqus2011}.

\subsection{Algorithm\label{sec:algorithm}}
The consistency between the EOS and return mapping equations under compactive plastic yield is facilitated by identifying a 2-step or ``split'' strain increment for the iterative return mapping solution, where the volumetric part is associated with previous step and the deviatoric part with the current. The split increment in strain is identified,
\beq \label{eq:split}
\Delta \bs{\varepsilon}_{\textrm{split}} := \Delta \bs{\varepsilon}'_{(n+1)} + \Delta {e}_{v,(n)}\bs{1}/3.
\eeq
We will drop the ``split'' subscript for brevity going forward, since all subsequent incremental equations assume this time-split definition given in Eq.~\eqref{eq:split}. This split is associated with the procedure for computing the shock-induced temperature and pressure, where the density and internal energy at the beginning of the increment ($t_{(n)}$) is used to update temperature and in turn pressure (see Algorithm Box~\ref{box:1}). \blue{In essence, it follows from the choice to calculate the EOS based upon the (liquid-solid) density at the start of the time increment and update it and the total density at the end of the increment based upon the partition of elastic and plastic volume strains solved for in the return mapping equations (Algorithm Boxes~\ref{box:1}-\ref{box:Yp}).}

Consistency between the EOS and the return mapping equations is achieved by updating the solid-liquid phase density with the converged pressure and current temperature. The temperature is fixed over the increment (being computed from previous converged solid-liquid density and specific internal energy), and the adiabatic bulk modulus is used within the iterative (Newton-Raphson) iteration (Algorithm Box \ref{box:nr}), $K_{nr} := \partial_{\rho^{s+l}} p |_T$, when pores crush over an increment. Either the shear modulus or Poisson's ratio can be held constant in the algorithm (and one must be), and the other is computed according to Hooke's law. 

Evaluation of the stress state locus is performed to evaluate if yielding is on the dilative (Yp-surface) or compactive (MCC-surface) side of critical state (or at critical state). The model does not include dilation nor damage when yielding on the ``dilative'' side, for which case a (non-iterative) radial return algorithm is utilized (Algorithm Box \ref{box:Yp}) to return to the Yp surface on the deviatoric plane. For the compactive side, a Newton-Raphson procedure is used to solve for the plastic multiplier and return-mapping of the stress given a trial elastic increment. The evolution equations are thus identified from Eq.~\eqref{eq:evols} with respect to $\mathcal{F}_{MCC}$ to be,
\begin{subequations} \label{eq:devols}
\beq
 \bs{\varepsilon}'^e_{(n+1)} =  \bs{\varepsilon}'^e_{(tr)} - \Delta \lambda \frac{\sqrt{6}q_{(n+1)}} {M^2_{(n+1)}} \hat{\bs{n}} ,
\eeq
\beq
e^e_{v, (n+1)} = e^e_{v, (tr)}  - \Delta \lambda (2p_{(n+1)} -p_{c,(n+1)}),
\eeq
\beq \label{eq:z_update}
z_{(n+1)} = z_{(n)} + \Delta \lambda \left( \frac{2q_{(n+1)}^2}{M_{(n+1)}^3} \frac{X P_{c0} M_0}{p_{c, (n+1)}^2}  -p_{(n+1)}    \right).
\eeq
\end{subequations}
These lead to the convenient update equations for the stress invariants of the yield function in terms of the plastic multiplier,
\begin{subequations}
\beq
p_{(n+1)}  = \frac{ p_{(tr)} + K_{nr}\Delta \lambda p_{c, (n+1)} } { 1 +2K_{nr}\Delta \lambda},
\eeq
\beq
q_{(n+1)} = \frac{ q_{(tr)} } {1+6G\Delta \lambda /M^2_{(n+1)}}.
\eeq
\end{subequations}
Further details of the algorithm are provided in Algorithm Boxes \ref{box:1}-\ref{box:Yp}.

%ALGORITHM 1

\begin{algorithm*}\small
    \caption{EOS/pore-crush/strength/damage coupled Yp-Cap material model algorithm\label{box:1}}
    {
    \begin{enumerate}
        \item{Given strain increment $\Delta\bs{\varepsilon}_{(n+1)}$ and previously converged state, construct time-split increment.
         \begin{algorithmic}%[o]
             \State {$\Delta \bs{\varepsilon}_{\textrm{split}} = \Delta \bs{\varepsilon}'_{(n+1)} + \Delta {e}_{v,(n)}\bs{1}/3$} \Comment{Eq.~\eqref{eq:split}}
         \end{algorithmic}
         }
        \item{Compute temperature, adiabatic bulk modulus and trial pressure from EOS tables given solid-liquid density $\rho^{s+l}$ and internal energy $E$ at $t_{(n)}$.
            \begin{algorithmic}%[1]
                 \State {$  T \quad =   \widehat{\textrm{EOS}} (\rho^{{s+l}}_{(n)}  ,E_{(n)})   $}
                \State{$p_{(tr)} = \widehat{\textrm{EOS}} (\rho^{{s+l}}_{(n)}, T_{(n)})$}
                \State {$K_{nr} = \dd{p_{(tr)}}{\rho^{{s+l}}} \Big|_{T} $}
                                         %\vspace{2mm}
                    %
            \end{algorithmic}
            }
              \item{Compute trial deviatoric stress and ISV given previously converged state and deviatoric strain increment at $t_{(n+1)}$.
            \begin{algorithmic}%[1]
                 \If { Is constant $G$}        
                    \State {$ G_{nr}=G_0$}
                     \Else { Is constant $\nu$ (from Hooke's law)}
                       \State {$ G_{nr} = \frac{3K_{nr} (1-2\nu_0)}{2(1+\nu_0)}$}
                       \EndIf
               \State{$\bs{s}_{(tr)} = 2G_{nr}\bs{\varepsilon}'$}
                       \State {$ q_{(tr)} =  \sqrt{3/2} \norm{\bs{s}_{(tr)}} $}
                    \State{$\beta_{(tr)} = Hz_{(n)} + \beta_{max} (\exp{\omega z_{(n)}} -1 )$} \Comment{Eq.~\eqref{eq:beta}}
                                        %\vspace{2mm}
                    %
            \end{algorithmic}
            }
              \item{Check stress locus to evaluate active yield surface.
            \begin{algorithmic}%[1]
                 \State{$p_{cs} = p_{c,(n)}/2$}
                 \State {$q_{cs} = M_{(n)}p_{cs} $}
                 \If { $p_{(tr)} = p_{cs} $ And $q_{(tr)} >q_{cs} $}        
                     \State{$q_{(n+1)}=q_{cs} = -M_{(n)}p_{cs} $} 
                     \State{$p_{(n+1)} = p_{(tr)}= p_{cs}$}
                     \State {$\mathcal{F}_{(tr)}=0$}
                     \ElsIf { $p_{(tr)} > p_{cs} $}
                      \State{$\mathcal{F}_{(tr)}:= \hat{\mathcal{F}}_{\textrm{Yp}}\left({\bs{\sigma}}_{(tr)}\right)$}
                      \Else
                        \If{Hardened = True}
                            \State{$\mathcal{F}_{(tr)} := \hat{\mathcal{F}}_{\textrm{Yp}}\left({\bs{\sigma}}_{(tr)}\right)$}
                        \Else
                            \State{$\mathcal{F}_{(tr)}:= \hat{\mathcal{F}}_{\textrm{MCC}}\left({\bs{\sigma}}_{(tr)}, \beta(z_{(tr)})\right)$}
                        \EndIf
                       \EndIf
                                 %\vspace{2mm}
                    %
            \end{algorithmic}
            }

                  \item{Check yield condition and perform return mapping if necessary.   
             \begin{algorithmic}%[1]
                \If {$\mathcal{F}_{(tr)} \leq 0$}
                    \State Set $(\bullet)_{(n+1)} = (\bullet)_{(tr)}$
                    \State {Continue}
                 \ElsIf {$\mathcal{F}_{(tr)}= \hat{\mathcal{F}}_{\textrm{Yp}}$}
                 \State {Compute plastic multiplier and equivalent plastic strain from Yp radial return Algorithm 2. }
                 \ElsIf {$\mathcal{F}_{(tr)}= \hat{\mathcal{F}}_{\textrm{MCC}}$}
                 \State {Compute plastic multiplier and equivalent plastic strain from MCC CPP return Algorithm 3. }
                \EndIf
            \end{algorithmic}
           }
               %\vspace{2mm}
                 \item{Compute EOS consistent density for converged value of pressure and current temperature \& specific internal energy.
             \begin{algorithmic}
             \If {MCC CPP algorithm was performed for return mapping} 
            \State {$\rho^{s+l}_{(n+1)} =  \widehat{\textrm{EOS}} \left(p_{(n+1)}, T   \right)$}
            \Else
            \State {$\rho^{s+l}_{(n+1)} = \rho^{s+l}_{(n)}/(1+\Delta e^e_{v,(n+1)})$}
            \EndIf
            \State {$E_{(n+1)} = E_{(n)} + 1/2\left(\bs{\sigma}_{(n)} + \bs{\sigma}_{(n+1)}  \right){:}\Delta\bs{\varepsilon}$} \Comment{$E$ computed as mid-step value.}
             \end{algorithmic}
            }
    \end{enumerate}
    }
    \label{alg:1}
    
\end{algorithm*}
%END ALGORITHM 1

%ALGORITHM 2
\begin{algorithm*}\small
    \caption{{ MCC closest point projection return mapping equations. The iterative process is continued until convergence with respect to a given tolerance is achieved.\label{box:nr}}}
    {
    \begin{enumerate}
        \item{Initialize with trial values, where $\mu:=G_{nr}\dl$
            \begin{algorithmic}%[1]
                \State{ $\mu^{(0)} = 0 \implies \mu^{(0)}=0$}
                \State{$\rm{\bs{v}}^{(0)} = [\Delta \lambda^{(0)}, p_{(tr)}, q_{(tr)}, \beta_{(tr)}]$}
            \end{algorithmic}
             }
        \item{Iterate
            \begin{algorithmic}%[1]
                \For {$ k = 1,\texttt{MAX\_ITERATION}\,$}
                    \State{$\mathcal{R}(\rm{\bs{v}}^{(k)}) := [R_{1}, R_{2}, R_{3}, R_{4}] $} \Comment{Used Eq.~\eqref{eq:residual}}
                    %\vspace{2mm}
                    \If {$\left|\mathcal{R}(\rm{\bs{v}}^{(k)})\right| \le \texttt{TOLERANCE}$}
                        \State \Return
                    \EndIf
                    %\vspace{2mm}
                    \State{$D\mathcal{R}(\rm{v}^{(k)}):= \begin{Bmatrix}
                    \frac{\partial R_{1}}{\partial\Delta\lambda} & \frac{\partial R_{1}}{\partial p} & \frac{\partial R_{1}}{\partial q} & \frac{\partial R_{1}}{\partial\beta} \\[2mm]
                    \frac{\partial R_{2}}{\partial\Delta\lambda} & \frac{\partial R_{2}}{\partial p} & \frac{\partial R_{2}}{\partial q} & \frac{\partial R_{2}}{\partial\beta} \\[2mm]
                    \frac{\partial R_{3}}{\partial\Delta\lambda} & \frac{\partial R_{3}}{\partial p} & \frac{\partial R_{3}}{\partial q} & \frac{\partial R_{3}}{\partial\beta} \\[2mm]
                    \frac{\partial R_{4}}{\partial\Delta\lambda} & \frac{\partial R_{4}}{\partial p} & \frac{\partial R_{4}}{\partial q} & \frac{\partial R_{4}}{\partial\beta} \\[2mm]
                    \end{Bmatrix}$}\Comment{Build Jacobian using Eqs. ~\eqref{eq:jacobian_start}-~\eqref{eq:jacobian_end}}
                   % \vspace{2mm}
                    \State{$\rm{\bs{v}}^{(k+1)} = \rm{\bs{v}}^{(k)} - D\mathcal{R}(\rm{\bs{v}}^{(k)})^{-1}\mathcal{R}(\rm{\bs{v}}^{(k)})$}\Comment{Update the value of $\rm{\bs{v}}$}
                    \State{$p_{c,(n+1)} = p_{c0} - \beta$}
                    \State{$M = M_{0}\left[1 - X + \frac{Xpc_{0}}{p_{c}^{2}}\right]$}\Comment{Update $M$ and $p_{c}$ for next iteration.}
                    \State{$z_{(n+1)} = \hat{z}(\mu, p_c, p,q)$} \Comment{Eq.~\eqref{eq:z_update}}
                   % \vspace{2mm}
                \EndFor
            \end{algorithmic}
            }
        \item { Update from converged $\rm{\bs{v}}$
            \begin{algorithmic}
                \State{$p_{cs} = \frac{ p_{c,(n+1)}}{2}$, $q_{cs} = -M_{(n+1)}p_{cs}$}
                \State{$\alpha_{(n+1)} = \frac{q_{cs}}{1 - R\exp{\frac{p_{cs}}{P_{y}}}}$}
                \State{$\gamma_{(n+1)} = R\alpha_{(n+1)}$}
                \State{$\bs{s}_{(n+1)} = \sqrt{\frac{2}{3}} q_{(n+1)}\hat{\bs{n}}_{(n+1)}$}\Comment{$\hat{\bs{n}}_{(n+1)} = \hat{\bs{n}}_{(tr)} =\bs{s}_{(tr)}/\norm{\bs{s}_{(tr)}}$}
            \end{algorithmic}
        \item {Check if air-filled pores have crushed-out (``hardened'')
            \begin{algorithmic}
                \If {$ z_{(n+1)} \ge z_{max}$}
                    \State{Hardened = True}
                \EndIf
            \end{algorithmic}
        }
        }    
    \end{enumerate}
    }
    \label{alg:plasticmultiplier_isoC}
\end{algorithm*} 
%END ALGORITHM 2

%ALGORITHM 3 Yp Radial Return
\begin{algorithm*}\small
    \caption{Yp Radial Return Algorithm. The mean stress $p$ is checked for exceeding the apex (if less than or equal to $a$). Then the Mises stress is returned to the yield surface given by the current values of $\alpha$ and $\gamma$. \label{box:Yp}}
    {
    \begin{enumerate}
        \item {Calculate
            \begin{algorithmic}
                \State{$a = P_{y}\log( \frac{\alpha}{\gamma}) $}
            \end{algorithmic}}
        \item {Compute $q_{(n+1)}$ and $p_{(n+1)}$ 
            \begin{algorithmic}
                \If{$p_{(tr)} \geq a$}
                    \State{$p_{(n+1)} = a$}
                    \State{$q_{(n+1)} = 0$}
                \Else
                    \State{$p_{(n+1)} = p_{(tr)}$}
                    \State{$q_{(n+1)} = \alpha - \gamma \exp{\frac{p_{(n+1)}}{P_{y}}}$}                
                \EndIf
            \end{algorithmic}}
        \item {Update deviatoric stress.
            \begin{algorithmic}
                \State{$\bs{s}_{(n+1)} = \sqrt{\frac{2}{3}} q_{(n+1)}\hat{\bs{n}}_{(n+1)}$} \Comment{$\hat{\bs{n}}_{(n+1)} = \hat{\bs{n}}_{(tr)} =\bs{s}_{(tr)}/\norm{\bs{s}_{(tr)}}$}
            \end{algorithmic}
        }
    \end{enumerate}
    }
    \end{algorithm*}
%END ALGORITHM 3

\subsection{Considerations for large strain\label{sec:large-def}}

The model as presented is extensible to large strains in either a hyperelastoplastic or hypoelastoplastic formulation in standard ways. We utilize a hypoelastoplastic formulation in the work presented here, for which we have found the results typically closely resemble the hyperelastoplastic one at least for ground-shock loading \citep[e.g.,][]{stahl2022probing}. The algorithm is straightforwardly extensible also to hyperelastoplasticity if isotropy of the material response is maintained. To model anisotropic elastoplasticity, further considerations are necessary that are not covered here \citep[see][]{bennett2019anisotropic}. We give a brief summary of large strain requirements in what follows, but refer to \citep[among others]{truesdell2004non, mourad2014incrementally} for further details of incrementally objective integration algorithms and objective stress-rates and to \citep[among others]{Bennettetal16,bennett2019anisotropic} for further details on large deformation hyperelastoplastic formulations for geomaterials. Extension of the current model to anisotropic large deformation hyperelastoplasticity is part of our intended future work.

For an incrementally objective constitutive formulation, an objective stress rate must be identified such that
\beq
\accentset{\circ}{\bs{S}} = \mathbb{C}{:}[\bs{d} - \bs{d}^p],
\eeq
where $\accentset{\circ}{\bs{S}}$ is an objective stress rate (typically a function of Cauchy stress, e.g., the Jaumann rate), $\mathbb{C}$ describes the constitutive response and the deformation rate $\bs{d} = (\bs{l} + \bs{l}^T)/2$ is the symmetric part of the velocity gradient $\bs{l}$ assumed additively composed of elastic and plastic parts, $\bs{d} = \bs{d}^e + \bs{d}^p$.
The hypoelastic stress constitutive relation thus cannot be directly derived from the elastic strain energy density in the Helmholtz free energy potential as described; however, we note that the potential does give the form of the constitutive relations and approaches it exactly if rotations are small (see \citep{Abaqus2011}).

For the stress to be hyperelastic, i.e., directly derived from the Helmholtz free energy potential, the constitutive relation must be in terms of the total stress and strains (not their rates), which requires essentially tracking of the deformation gradient, $\bs{F}$, for its use in constructing appropriate strain measures and mapping stress measures \citep[cf.,][]{truesdell1960classical, Bennettetal16}. For isotropy, the logarithmic strain (a.k.a. Hencky strain), $\ln{\bs{v}}$, where stretch $\bs{v}$ and rotation $\bs{R}$ are formed from the polar decomposition $\bs{F}=\bs{v}\bs{R}$ and Cauchy stress scaled by the determinant of the deformation gradient, the Kirchhoff stress, allow for a formulation that essentially follows that for small strain \citep{simo1992algorithms}.

%SECTION 5
\section{Example results and discussion\label{sec:examples}}

We provide two sets of example simulations: (1) material point simulations and (2) simulations of a large underground chemical explosion with parameter calibration. Calibration to the underground explosion data is subsequently discussed in Section \ref{sec:FE}. The material point simulation utilizes these calibrated parameters, but with a much larger air-filled porosity of 20\% to demonstrate pore-crush effects. Material parameters for both sets of simulations are shown in Table~\ref{tab:params}.

\begin{table}
\begin{centering}
\centering
\par\end{centering}
\centering{}\caption{Yp-Cap model parameters calibrated to the NPE and altered for material point simulations.\label{tab:params}}
\small%
\begin{tabular}{llcccc}
\hline 
\multirow{2}{*}{Physical parameter} & \multirow{2}{*}{Symbol} & \multirow{2}{*}{Material point} & \multirow{2}{*}{NPE} & \multirow{2}{*}{Units} & \multirow{2}{*}{Calibration procedure}\tabularnewline
 \tabularnewline
\hline 
\hline \\[-2.3ex]
Sesame table tuff, 		          &$\#$ 				& 7122     & 7122			& -		& Chosen \tabularnewline
Initial density, 				  &$\rho_0$ 			& 1608.80  &1910.25		& kg/m$^3$		 & Measured \tabularnewline
Initial temperature, 			  &$T_0$ 				& 298.15   &298.15		& Kelvin		  & Approximated \tabularnewline
Initial shear modulus, 		      &$G_0$ 				& 4.557    &3.972		& GPa	   &  (Shear wave speed) Measured\tabularnewline
Linear hardening coeff. , 		  &$H$ 				    & 1.0      &7.0		&GPa		& Crush-curve\tabularnewline
Exponential hardening coeff., 	  &$\beta_{\textrm{max}}$&0.50 	   &2.0		& MPa		& Crush-curve \tabularnewline
Exponential hardening coeff., 	  &$\omega$ 			& 50.0	   &255.0	& -		& Crush-curve \tabularnewline
Initial preconsolidation stress,  &$p_{c0}$ 			& -950.0     &-95.0  & MPa		& Crush-curve \tabularnewline
Damage propensity factor, 	      &$X$ 				    & 0,1		&1		& -		& Velocity gauges\tabularnewline
Yp initial cut-off stress,        &$\alpha$             & 0.6396     &0.1617       & GPa      & Kamm \& Bos\tabularnewline
Yp initial parameter, 	          &$\gamma$ 		    & 0.6396	 &0.1436	&GPa 		& Kamm \& Bos\tabularnewline
Yp exponential form parameter,    &$P_y$ 				& 0.35		&0.175		& GPa		& Kamm \& Bos \tabularnewline
Maximum crushable porosity, 	  &$z_{\textrm{max}}$ 	& 0.20		&	0.0217	& -		& Measured \tabularnewline\\[-2.3ex]
\hline 
\end{tabular}
\end{table}

\subsection{Material point simulations\label{sec:mat-point}}

\begin{figure}[htbp]
\centering
\begin{subfigure}[b]{0.45\textwidth}
\includegraphics[width=\textwidth]{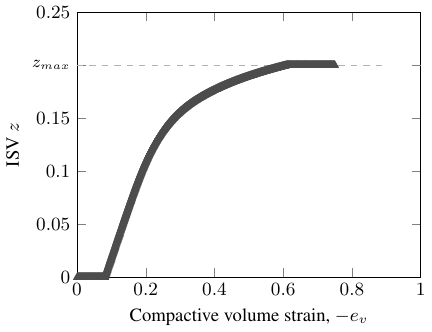}%main-figure6.pdf}
\caption{Air-filled porosity crush-out}
\end{subfigure}
\hfill
\begin{subfigure}[b]{0.45\textwidth}
\includegraphics[width=\textwidth]{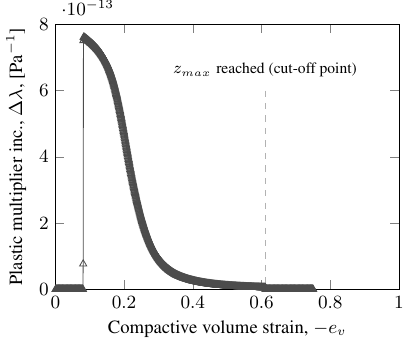}%main-figure7.pdf}
\caption{\blue{Cut-off of crush at negligible plastic volume strain}}
\end{subfigure}
\caption{\blue{Crush out of ISV $z$ through exponential hardening. Crush is cut-off at $z_{max}$ (a), which corresponds to relatively very-small increments of plastic multiplier (b).}}
\label{fig:z}
\end{figure}

\begin{figure}[htbp]
\centering
\begin{subfigure}[b]{0.45\textwidth}
\includegraphics[width=0.94\textwidth]{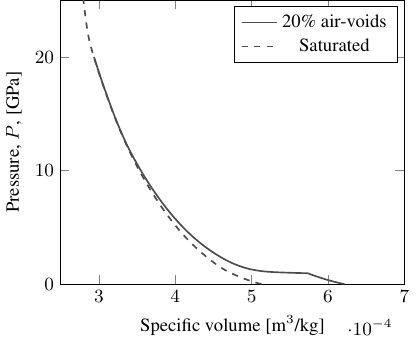}%main-figure8.pdf}
\caption{Comparison of saturated tuff versus 20\% air-filled voids.}
\label{fig:specificvolume}
\end{subfigure}
\begin{subfigure}[b]{0.45\textwidth}
\includegraphics[width=\textwidth]{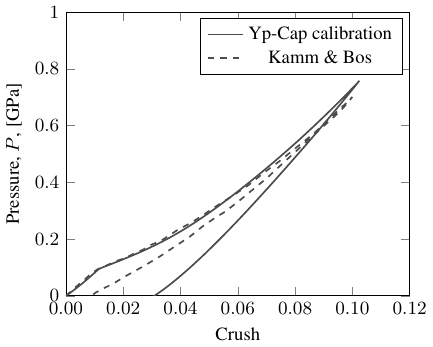}%main-figure9.pdf}
\caption{Crush curve for tuff at NPE working point.}
\label{fig:crushcurve}
\end{subfigure}
\caption{Hydrostatic compression using tuff Sesame Table 7221. Loading curves show combined pore-crush and EOS response, and unload curve is EOS response without crush. (a) 20\% air-filled porosity crushed out at high pressure vs. specific volume ($=1/\rho$). (b) calibration to crush ($\rho/\rho_0 -1$) curve reported for NPE working point by \citet{kamm1995comparison}.}
\label{fig:crushcurves}
\end{figure}

Numerical experiments were conducted to exemplify the the model results for a partially saturated tuff. The 7122 Sesame Table \citep{johnsonSesame} for saturated tuff was used, which consists of 62.0\% SiO$_2$, 13.5\% Al$_2$O$_3$, 24.5\% H$_2$O by weight. Partial saturation was achieved by calibrating the crush potential to crush-out a 20\% air-filled void volume fraction (see Fig.~\ref{fig:z}), which corresponds to degree of saturation, $S=\phi^l/(\phi^l + \phi^g)$, of approximately 62\%. Fig.~\ref{fig:specificvolume} compares the pressure-volume relationship for the $S=$62\% tuff vs.~$S=100\%$ tuff, i.e., absent the air-filled voids volume fraction but with the water-filled void volume fraction. A second crush calibration was performed for 2\% air filled porosity for the calibration to the NPE experiment discussed in the next section, shown in Fig.~\ref{fig:crushcurve}.

The pore-crush potential $\Psi^s$ of Eq.~\eqref{eq:storedenergy} describes an exponential crush-out of porosity (or a logarithmic pressure vs.~volume relationship), where the change to the ISV $z$ is equal to change in porosity on the hydrostatic axis (when $q=0$), which can be seen by solving Eq.~\eqref{eq:evols} to see $\Delta z = -\Delta \lambda p$ and $\Delta e^p_v |_{q=0} = -\Delta z$. To be physically meaningful up to high-pressure loading, the evolution of $z$ must exhibit crush-out. With proper selection of parameters, the functional form of $\Psi^s$ does, which is evident in Fig.~\ref{fig:z}. A further consideration is however necessary. \blue{As $\Delta z$ approaches zero (with complete crush-out of porosity), the increments of plastic volume strain also approach zero; thus iterating for the partition of elastic and plastic strains may lead to machine precision issues and is an additional computational cost that does not significantly affect the model results. Although the Newton-Raphson iteration we employ is robust (see the Appendix), we avoid these potential pitfalls by pre-calibrating as a parameter a cut-off value $z$ at which no further pore-crush is possible, namely the air-filled void volume fraction. In practice, we find any artifact of the cut-off to be negligible and that it ensures robust convergence of the iterative return-mapping equations if properly calibrated (such as exemplified in Fig.~\ref{fig:z}b).}

%FIGURE

%FIGURE

\begin{figure}[htbp]
\centering
\begin{subfigure}[b]{0.45\textwidth}
\includegraphics[width=\textwidth]{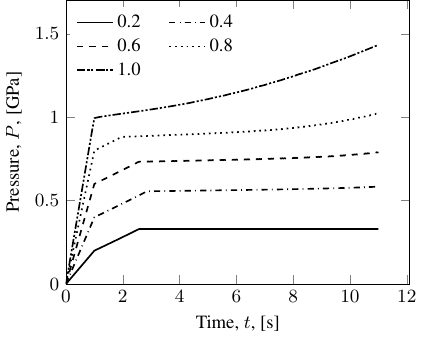}%main-figure10.pdf}
\caption{$X=0$}
\end{subfigure}
\hfill
\begin{subfigure}[b]{0.45\textwidth}
\includegraphics[width=\textwidth]{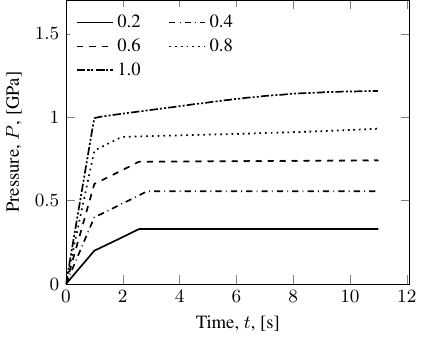}%main-figure11.pdf}
\caption{$X=1$}
\end{subfigure}
\caption{Pressure history for various confining pressures of numerical triaxial tests. Compression at a rate of 0.2 m/s begins at $t=1$.}
\label{fig:tcp}
\end{figure}

\begin{figure}[htbp]
\centering
\begin{subfigure}[b]{0.45\textwidth}
\includegraphics[width=\textwidth]{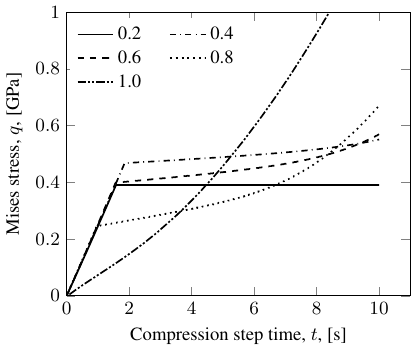}%main-figure12.pdf}
\caption{$X=0$}
\end{subfigure}
\hfill
\begin{subfigure}[b]{0.45\textwidth}
\includegraphics[width=\textwidth]{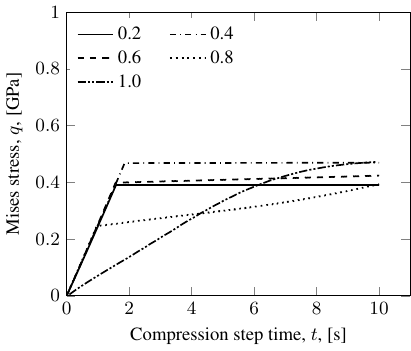}%main-figure13.pdf}
\caption{$X=1$}
\end{subfigure}
\caption{Yield and crush-hardening under triaxial compression. Compression step time (at a rate of 0.2 m/s) begins at $t=0$.}
\label{fig:tcq}
\end{figure}

To exemplify the Yp-Cap yield surface and its evolution under crush, numerical Triaxial Compression (TC) tests were performed at various confining pressures (Figs.~\ref{fig:tcp}-\ref{fig:tcq}). Yp yield surface parameters were chosen to recover values of $M_0=R=1$. Comparison between pore-crush damage and no-damage, i.e., values of $X=0$ and $X=1$, are provided in the figures. The effect of the parameter $X$ on hardening under pore-crush is exemplified, where the limiting of peak strength with $X=1$ is evident in the high confining pressure curves. No hardening (``perfect plasticity'') is evident at confining pressures dilative of CS, with hardening on the compactive side as expected.

%FIG TCP

%FIG TCQ

\begin{figure}[htbp]
\centering
\begin{subfigure}[b]{0.495\textwidth}
\includegraphics[width=\textwidth]{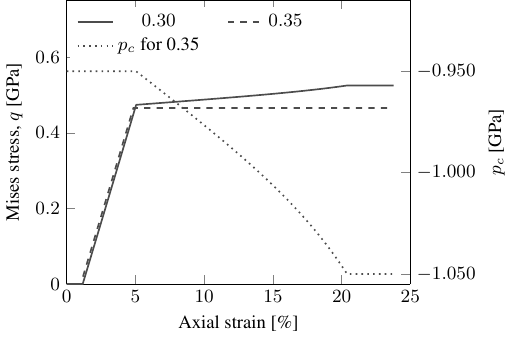}%main-figure12.pdf}
\caption{$X=0$}
\end{subfigure}
\hfill
\begin{subfigure}[b]{0.495\textwidth}
\includegraphics[width=\textwidth]{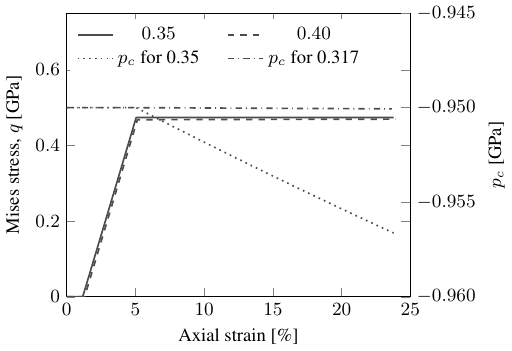}%main-figure13.pdf}
\caption{$X=1$}
\end{subfigure}
\caption{\blue{Yielding near Yp and MCC surface intercept (brittle-ductile transition point).}}
\label{fig:cs}
\end{figure}

\blue{Figure~\ref{fig:cs} shows Triaxial Compression (TC) load paths intersecting the yield surface near the transition point from the Yp to MCC surfaces. Figure \ref{fig:cs}a shows confining pressures that put the yield point on either side of intersection (the so-called critical state or also brittle-ductile transition) for $X=0$. For this case, the load path intersecting near the transition point on the MCC side (0.35 GPa confining pressure) hardens until the stress state coincides with the transition point, at which time it becomes perfectly plastic. This is also evident by the evolution of the $p_c$ in the figure. Any TC load path that intersects the transition point or any point on the Yp surface, such as the 0.30 GPa confining pressure shown, (for any value of $X$) becomes perfectly plastic in yielding. This is because of the split time increment and the radial return algorithm. Notably, the return path is smooth in transition across the Yp and MCC intersection point, being vertical in Fig.~\ref{fig:evolsurf} on the Yp side and also at the intersect with the MCC surface, which is horizontal in Fig.~\ref{fig:evolsurf} (with normal return). }

\blue{Figure~\ref{fig:cs}b shows load paths near the transition point on the MCC side for $X=1$. A confining pressure of 0.317 GPa puts the yield point at nearly exactly at the yield surface transition point, where it can be seen $p_c$ does not evolve and the perfect plasticity ensues. Slightly further away on the MCC side (the 0.35 GPa confing pressure), $p_c$ evolves but considerably less than it does for $X=0$. As the yield stress moves further away from the intersection, the hardening increases, evident by comparing 0.35 and 0.40 GPa confining pressures in Fig.~\ref{fig:cs}b.
}

\begin{figure}[htbp]
\centering
\begin{subfigure}[b]{0.40\textwidth}
\includegraphics[width=\textwidth]{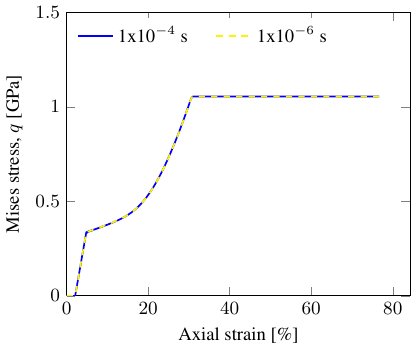}%main-figure12.pdf}
\caption{$X=0$}
\end{subfigure}
\hfill
\begin{subfigure}[b]{0.40\textwidth}
\includegraphics[width=\textwidth]{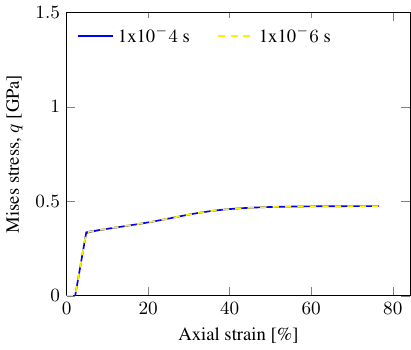}%main-figure13.pdf}
\caption{$X=1$}
\end{subfigure}
\caption{\blue{Results of time step size study showing no influence on pressure dependent strength from time step sizes ranging from 1x10$^{-4}$ to 1x10$^{-6}$ seconds (plots lie on top of each other). Compression loading was at a constant acceleration of 10 m/s$^2$ run to large strain (nearly 80\% axial strain) after first loading to a confining pressure of 700 MPa (that intercepted approximately in the middle of the MCC cap surface). Results also reinforce results in Fig.~\ref{fig:cs} showing smooth transition to perfect plasticity at critical stress state.}}
\label{fig:tstep}
\end{figure}

\blue{A study was performed to evaluate if the model is sensitive to time increment size due to the split time step assumption of Eq.~\eqref{eq:split}. The results are exemplified in Figure \ref{fig:tstep}, showing no influence of time step size on pressure dependent strength results. A confining pressure of 700 MPa was chosen to intersect the MCC cap surface approximately in its middle, where plastic volumetric and deviatoric strains are at first similar in magnitude. A constant acceleration was applied for the compression step of 10 m/s$^2$ for 1 second, resulting in large deformation. Even with constantly increasing velocity and run to large strains, no influence of time step size was discernible. This gives confidence that the split time step approach to coupling the EOS with pressure dependent strength and crush is a reasonable approximation, though we note qualitative evaluation of error isn't possible since we don't have a model (or presently know how to construct such a model) that evaluates pressure through an EOS also at the end of the increment used in pressure dependent return mapping equations with volumetric plasticity. However, it should also be noted that the linearization of pressure during the iterative solution for partitioning of elastic and plastic strains and subsequent update to density state is consistent with all thermodynamic and constitutive equations put forward in Sections \ref{sec:methods} and \ref{sec:theory}. This, in combination with the fact that the model is not sensitive to time increment size, gives us confidence in the model results and the method of utilizing a split time increment for the volumetric/deviatoric stress solution coupling.
}

%FIG CS

%FIG TSTEP

\subsection{Finite Element simulation of the NPE\label{sec:FE}}

Calibration of the model is conducted to the Non-Proliferation Experiment (NPE), an underground explosion conducted in tuff at the Nevada Nuclear Security Site (formerly called the Nevada Test Site) in September of 1993 \citep{denny1994proceedings, kamm1995comparison}. The NPE source consisted of 1.29$\times 10^6$ kg of ANFO-emulsion blasting agent, with an estimated TNT equivalent energy of 1.07 kiloton. The NPE was chosen for an example application of the model because significant openly available documentation was provided by \citet{kamm1995comparison}, for which our model shares the same exponential form of pressure-dependent strength absent the pore-crush (the Yp surface), allowing insightful comparison with their results. \blue{We note that our intent here is not to perform further analysis of the NPE, which Kamm and Bos have already done, but rather to demonstrate that the model framework presented here is capable of modeling the ground shock generated by the NPE while integrating an EOS with pressure-dependent strength, pore-crush and damage. The calibration is further performed to demonstrate the robustness of the model in finite element simulations and to demonstrate the effect of including pore-crush and associated strength-degradation in the model results (versus neglecting those phenomena).} 

\begin{figure}
\centering
\includegraphics[width=\textwidth]{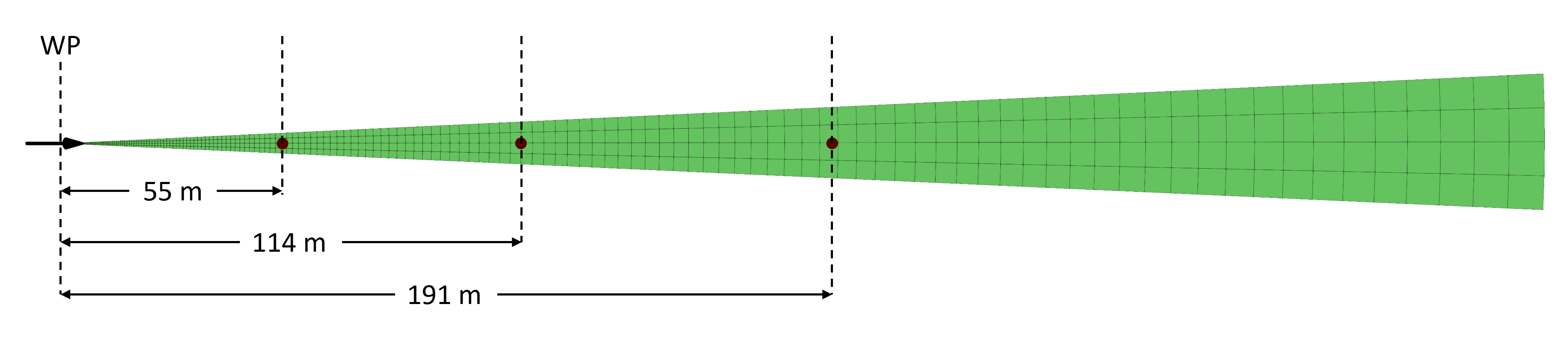}
\caption{3D spherical symmetry (isotropy) modeled with 3D wedge, pressure loading at cavity wall, mesh, and sensor locations corresponding to field gauge locations of the NPE.\label{fig:wedgemesh}}
\end{figure}

The model is conducive to a straightforward sequential calibration of parameters as follows: (1) EOS, (2) crush-curve, (3) strength, (4) tuning. The EOS parameters in this case are taken again as the Sesame table 7122 for saturated tuff \citep{sesame}, but we emphasize that any other EOS parameterization can be used with the model. The crush properties, including hardening parameters $\{H,\beta_{\textrm{max}}, \omega\}$ and preconsolidation stress $p_{c0}$, are best calibrated to isotropic compression measurements in our view (because they isolate the pressure-crush response), but can also be calibrated to 1D compression (oedometer) or, lacking such data, triaxial compression measurements in conjunction with the strength parameters. For the NPE, we calibrated to the crush-curve of the tuff at the source placement (working point) reported by \citep{kamm1995comparison} (Fig.~\ref{fig:crushcurve}). The strength parameters ($R$ and $P_y$) are typically best calibrated to triaxial compression measurements, but for the NPE we were able to directly use those reported by \citep{kamm1995comparison}. Notably, the propensity for damage parameter $X$ typically requires some sort of repeated or cyclic testing, but lacking this type of data for the NPE, we incorporate into the last step, ``tuning.'' We call the final part of the calibration process tuning because this takes into account any other information that may be available. For the case of the NPE, this is macroscopic (field-scale) velocity measurements that characterize the material's dynamic response at specific distances from the source (Fig.~\ref{fig:wedgemesh}).

\begin{figure}[htbp]
\centering
\small
\includegraphics[width=0.40\textwidth]{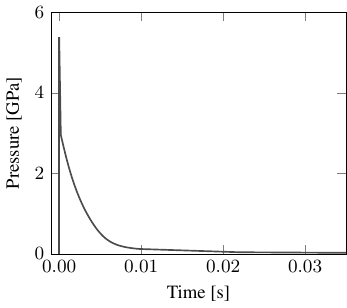}
\caption{Applied pressure loading at cavity wall was fit to previous hydrocode simulations of the chemical explosion. Peak pressure is 5.38 GPa.\label{fig:pressureloading}}
\end{figure}

\begin{figure}[htbp]
\centering
\begin{subfigure}[b]{0.497\textwidth}
\includegraphics[width=\textwidth, trim={3.2cm 1.7cm 3.2cm 1.1cm},clip]{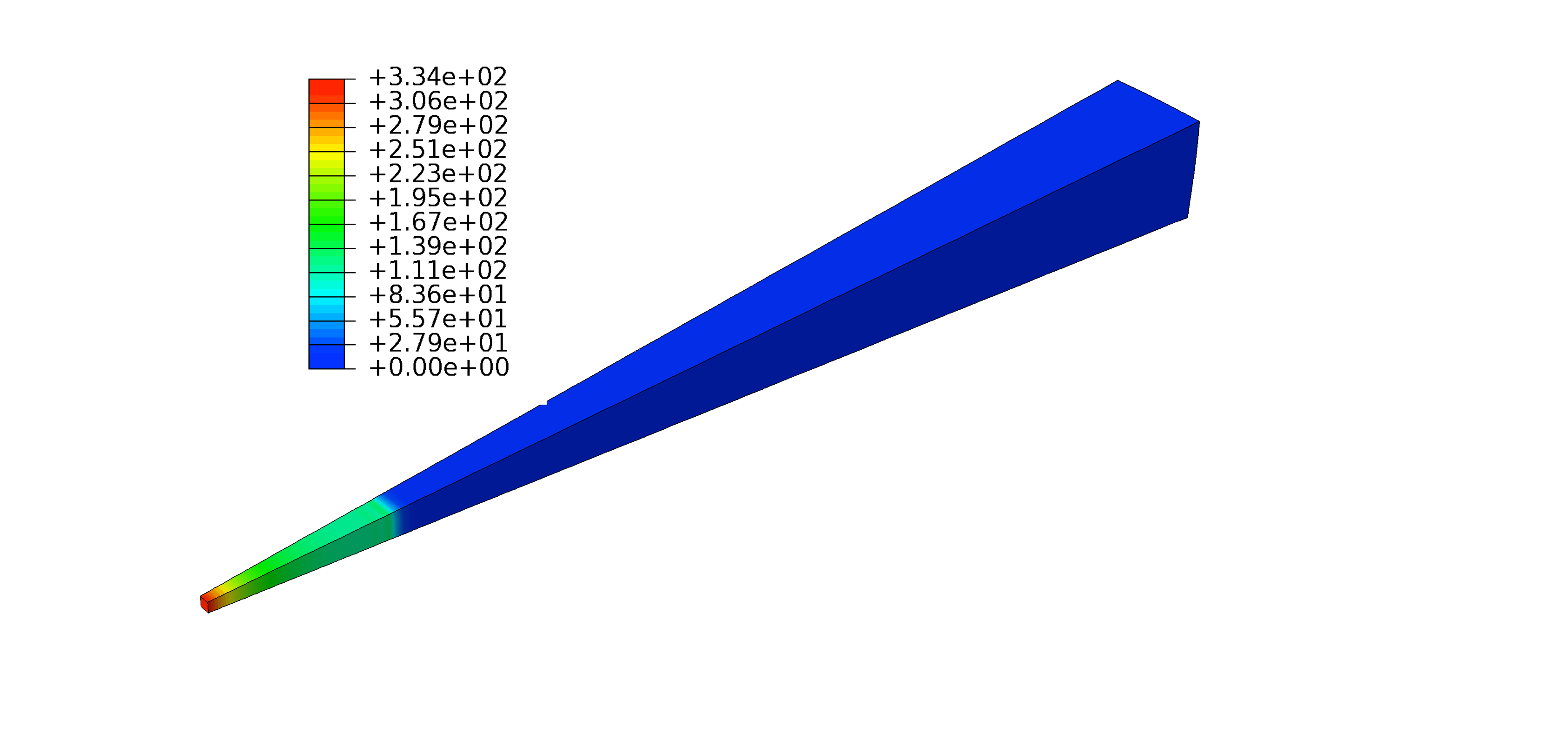}%velocity_prop_0s.png}
\caption{Radial velocity [m/s]}
\end{subfigure} %end subfigure
\hfill
\begin{subfigure}[b]{0.497\textwidth}
\includegraphics[width=\textwidth, trim={3.2cm 1.7cm 3.2cm 1.1cm},clip]{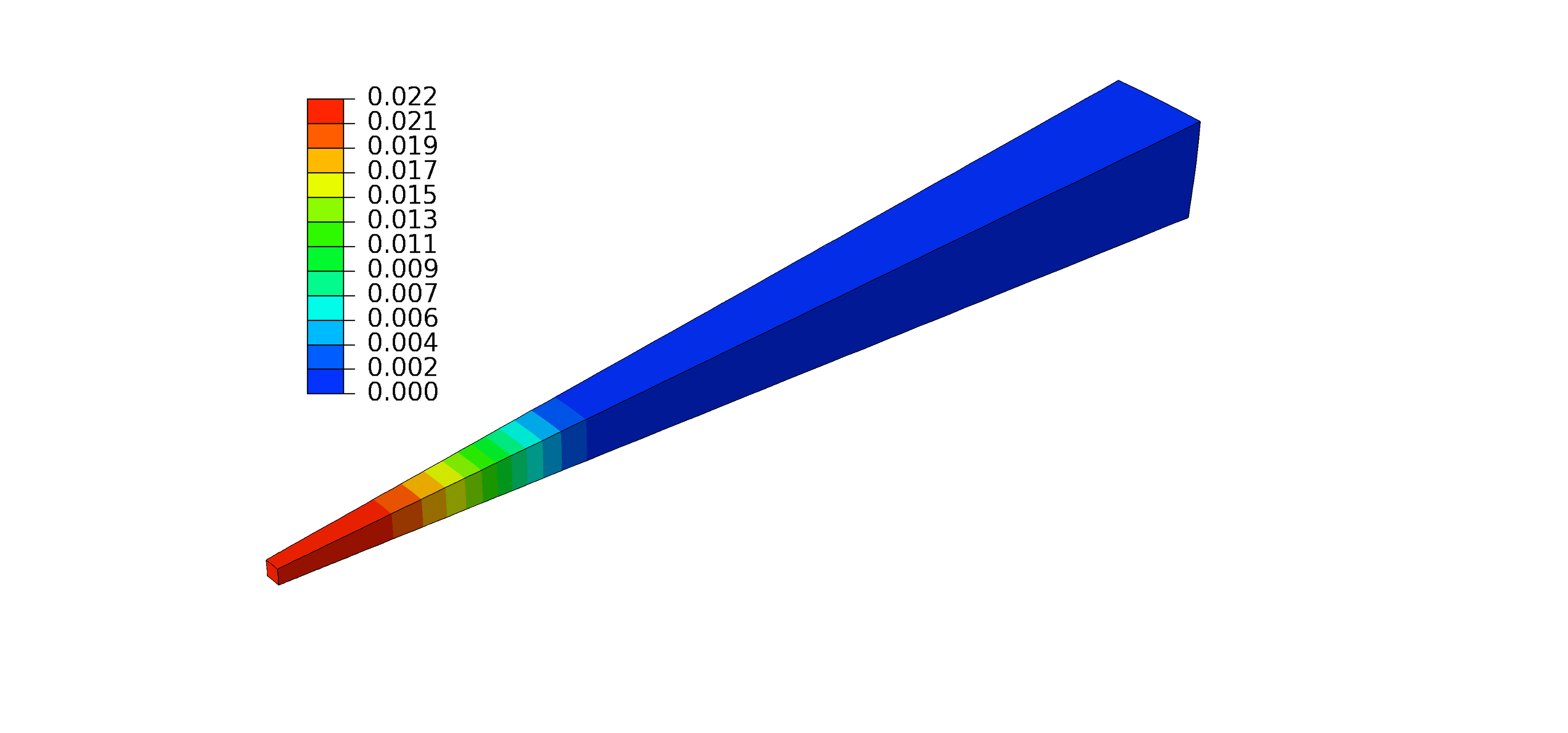}%z_final.png}
\caption{Crush out $z$}
\end{subfigure} %end subfigure
\caption{Velocity at the end of pressure loading (approximately 0.01 s) (a), and total pore-crush at end of simulation (b).}
\label{fig:wedgecrush}
\end{figure}

\begin{figure}[htbp]
\centering
\begin{subfigure}[b]{0.49\textwidth}
\includegraphics[width=\textwidth]{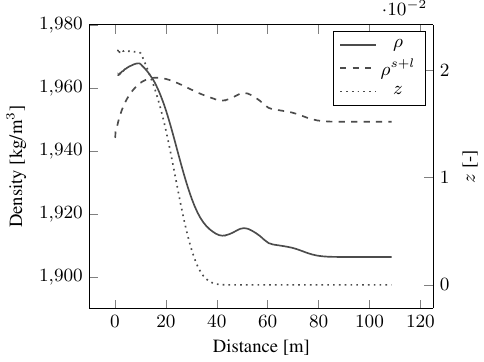}%press_prop_0s.png}
\caption{Density profile}
\end{subfigure} %end subfigure
\hfill
\begin{subfigure}[b]{0.49\textwidth}
\includegraphics[width=0.87\textwidth]{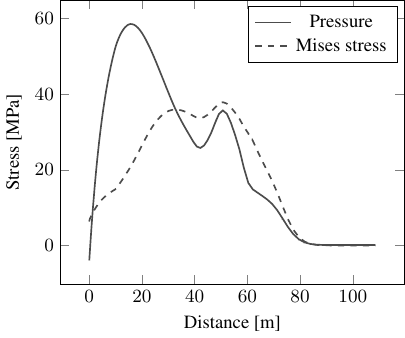}%mises_prop_0s.png}
\caption{Stress profile}
\end{subfigure} %end subfigure
\caption{\blue{Line plots of ground density and stress state with distance from cavity wall at the time peak velocity is observed at the 55 m gauge.}}
\label{fig:wedgestress0}
\end{figure}

\begin{figure}[htbp]
\centering
\begin{subfigure}[b]{0.49\textwidth}
\includegraphics[width=\textwidth]{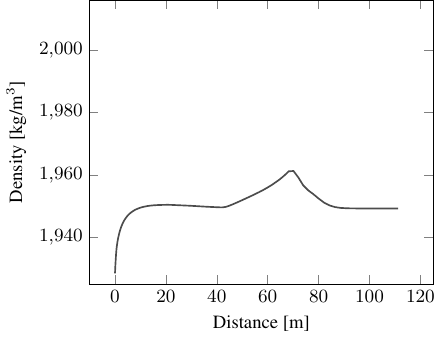}%press_prop_0s.png}
\caption{Density profile}
\end{subfigure} %end subfigure
\hfill
\begin{subfigure}[b]{0.49\textwidth}
\includegraphics[width=0.937\textwidth]{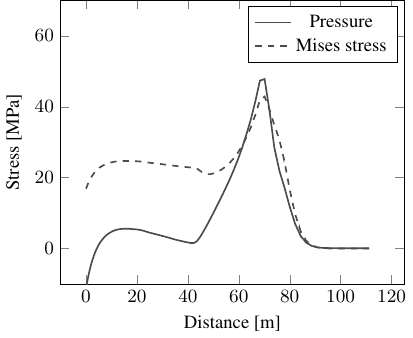}%mises_prop_0s.png}
\caption{Stress profile}
\end{subfigure} %end subfigure
\caption{\blue{For comparison, results with pore-crush suppressed of ground density and stress state at the time peak velocity is observed at the 55 m gauge.}}
\label{fig:wedgeNoCrush}
\end{figure}

\begin{figure}[htbp]
\centering
\begin{subfigure}[b]{0.49\textwidth}
\includegraphics[width=\textwidth]{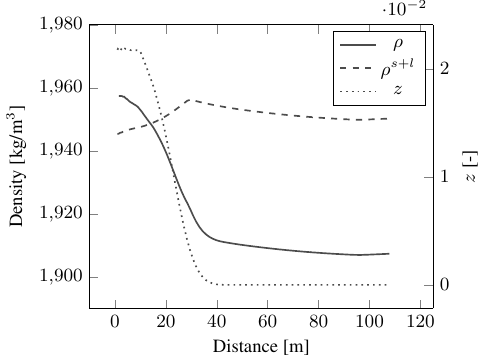}%press_prop_0s.png}
\caption{Density profile}
\end{subfigure} %end subfigure
\hfill
\begin{subfigure}[b]{0.49\textwidth}
\includegraphics[width=0.879\textwidth]{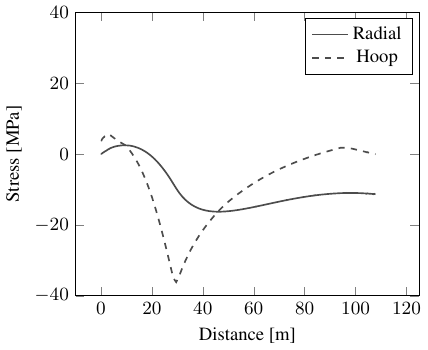}%mises_prop_0s.png}
\caption{Containment stress profile}
\end{subfigure} %end subfigure
\caption{\blue{Line plots of residual ground density and stress state with distance from cavity wall when velocity is observed to cease in the first 100 m (at 0.2175 s). Final cavity radius is 17.23 meters, which compares well to Kamm and Bos estimates between 16.5 and 19 m.}}
\label{fig:wedgeResidual}
\end{figure}

\begin{figure}[htbp]
\centering
\small
\begin{subfigure}[b]{0.32\textwidth}
\includegraphics[width=\textwidth]{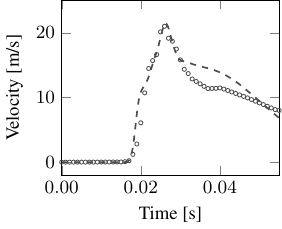}
\caption{55 m gauge}
\end{subfigure} %end subfigure
\hfill
\begin{subfigure}[b]{0.32\textwidth}
\includegraphics[width=\textwidth, height=0.85\textwidth]{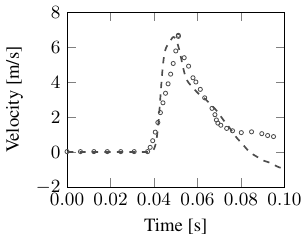}
\caption{114 m gauge}
\end{subfigure}
\hfill
\begin{subfigure}[b]{0.32\textwidth}
\includegraphics[width=\textwidth]{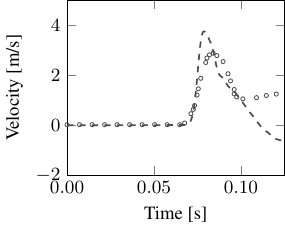}
\caption{191 m gauge}
\end{subfigure}
\caption{Radial velocity shock wave profiles of 3D wedge (dashed lines). Comparison to \citet{kamm1995comparison} reported gauge measurements (open circles).}
\label{fig:bestfit_gauge}
\end{figure}

The pressure loading from the explosive source at the cavity wall was used to simulate the explosion (Fig.~\ref{fig:pressureloading}). This was obtained from an exponential fit to the pressure wave simulated in previous source simulations of the NPE chemical explosive source at Los Alamos National Laboratory. \blue{Figures~\ref{fig:wedgecrush}-\ref{fig:wedgeResidual} show the simulation results. Figure~\ref{fig:wedgecrush}a shows the radial velocity with distance from the source at the approximate end of the pressure loading impulse, $t=0.01$ s. The total amount of pore crush-out is shown in Fig.~\ref{fig:wedgecrush}, which shows a large region of complete porosity crush out diminishing with distance to eventual no-crush, as might be expected. Figure~\ref{fig:wedgestress0} shows the density and stress fields from the cavity wall to 100 m radially away at approximately the point in time (0.027 s) that the peak velocity reaches the 55 m gauge evident in Fig.~\ref{fig:bestfit_gauge}. The stress field interestingly exhibits a double peak shock structure, perhaps resulting from crush-hardening. The stress field is considerably changed by the inclusion of pore-crush in the model, which is evident by comparing Figs.~\ref{fig:wedgestress0} and Fig.~\ref{fig:wedgeNoCrush}. When pore-crush is suppressed in the model (Fig.~\ref{fig:wedgeNoCrush}) even for the air-filled porosity of only 2.2\%, the stress peak and profile is markedly different. Figure~\ref{fig:wedgeResidual} shows the predicted residual state of the density and stress fields when ground motion ceases, at approximately 0.2175 s. Stress is plotted in radial and hoop stresses, a convention for evaluating ``containment'' stress field \citep{terhune1977containment, terhune1978analysis}, which is where the compressive hoop stress magnitude exceeds that of the radial stress (predicted out to about 45 m in the model). The model predicted a final cavity radius of 17.23 meters (growth from the initial radius of 6.45 m), which is consistent with that reported in Kamm and Bos of between 16.5 and 19 m.
}

%FIGURE

%FIGURE PRESSURE LOADING

%FIGURE WEDGE 

%FIGURE LINE PLOTS

%FIGURE

%FIGURE

\begin{figure}[htbp]
\centering
%\small
\begin{subfigure}[b]{0.32\textwidth}
\includegraphics[width=\textwidth, height=0.80\textwidth]{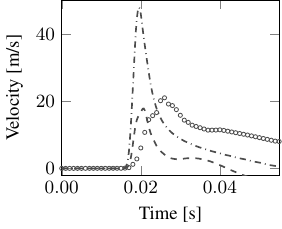}
\caption{55 m gauge}
\end{subfigure} %end subfigure
\hfill
\begin{subfigure}[b]{0.32\textwidth}
\includegraphics[width=\textwidth, height=0.81\textwidth]{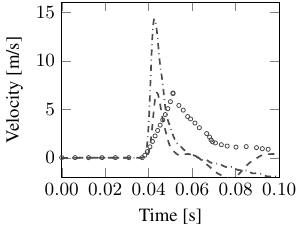}
\caption{114 m gauge}
\end{subfigure}
\hfill
\begin{subfigure}[b]{0.32\textwidth}
\includegraphics[width=\textwidth, height=0.84\textwidth]{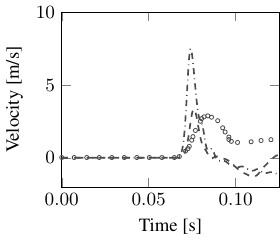}
\caption{191 m gauge}
\end{subfigure}
\caption{Radial velocity of 3D wedge. Open circles are gauge measurements; dashed lines are $X=0$; dashed-dotted lines are without pore-crush (Yp surface without Cap).}
\label{fig:x0comp}
\end{figure}

The best fit to the velocity gauge measurements is presented in Fig.~\ref{fig:bestfit_gauge}, which we feel is quite good considering the only further parameters ``tuned'' for this fit were $\{X,\alpha_0, \gamma_0\}$. The effect of not-including the cap surface (using just the ``Yp'' yield surface) and removing the propensity for pore-crush damage (setting $X=0$) is shown in Fig.\ref{fig:x0comp}. Notably, propensity for pore-crush $X\approx1$ was needed to match the gauge measured shock-profiles, where without damage, the modeled shock profile has a lower peak and could not be broadened to match the measurements through the other parameters. Removing the Cap-surface (removing pore-crush modeling), resulted in a similarly too sharp shock profile and also too high of a peak velocity.

%FIGURE SIGNAL COMPARISON 1 BEST FIT

%FIGURE SIGNAL COMPARISON 2

%SECTION 6
\section{Conclusion}
The novel ground-shock material model, ``Yp-Cap,'' for integrating EOS with pore-crush, strength and damage has been described. Numerical material point and FEM simulations of the NPE underground chemical explosive test have demonstrated both the physical basis of the model in coupling saturated geomaterial EOS to air-filled void crush and the effectiveness of the model in accurate simulation of ground-shock profiles. Mixture theory relating to the model assumptions is described, where discriminating between the density of the combined solid and liquid phases versus the total density including also air-filled void volume plays a critical role in EOS/pore-crush coupling. Crush-out of air-filled porosity is enabled through identifying an appropriate crush-potential that enters the Helmholtz free energy, providing a thermodynamicaly consistent description that couples Cap-plasticity with EOS response. A novel Modified Cam-Clay (MCC) type of surface is proposed and merged with a non-linear pressure dependent yield surface (the ``Yp'' surface) at the critical state (a.k.a.~ductile-brittle transition point for rocks). A novel ``propensity for damage'' parameter ($X$) is introduced into the dependence of the critical state line on the preconsilidation stress in order to recover the classical MCC surface at one extreme and to provide a tunability of strength-damage from pore-crush. The algorithm, including a split time increment, is described in detail in the interest of reproducibility. 

The model is employed in finite element simulations of the NPE with description of the suggested calibration procedure. Model results demonstrate the effectiveness of including the Cap-surface with a traditional non-linear pressure dependent yield surface (the ``Yp'' surface), where pore-crush is found necessary to simulate measured shock-profiles even at relatively low air-filled void fractions ($\approx$ 1\% - 2\%). It was also found that propensity for pore-crush damage was required to match the breadth of the measured shock profile, with comparison provided to no-damage model predictions.

\section{Acknowledgements}
This work was supported by the US Department of Energy through
the Los Alamos National Laboratory (LANL). LANL is operated by Triad National Security, LLC, for the National Nuclear
Security Administration of US Department of Energy (Contract No.
89233218CNA000001). This Low Yield Nuclear Monitoring (LYNM) research was funded by the National Nuclear Security Administration, Defense Nuclear Nonproliferation Research and Development (NNSA DNN R\&D). The authors acknowledge important interdisciplinary collaboration with scientists and engineers from LANL, LLNL, NNSS, PNNL, and SNL, as well as support from LANL's Laboratory Directed Research and Development (LDRD) program. LA-UR-23-34028.

\appendix \label{Appendix:A}

%\section{Derivation of MCC pore-crush damage yield surface\label{sec:MCCD}}

\section{Newton Raphson iteration for crush}

The MCC yield function is the only active surface during iteration. The independent variables are taken to be $x:=\{\mu, p, q, \beta\}$, where unitless $\mu:= G_{nr} \Delta \lambda$ and the residual is defined,
\beq \label{eq:residual}
R(\mu, p, q, \beta):= 
\begin{Bmatrix} 
\mathcal{F}_{MCC} (p,q,\beta) \\
p(1 + 2\mu K_{nr}/G_{nr} ) - p_{(tr)} - \mu (p_{c0} - \beta) K_{nr}/G_{nr}    \\
q(1+6\mu/M^2) - q_{(tr)} \\
\beta - H(z_{(n)} + \Delta z ) - \beta_{max} (\exp{\omega z_{(n)}} \exp{\omega \Delta z} -1 ) 
\end{Bmatrix}.
\eeq
where $\Delta z = \Delta \hat{z} (\Delta \lambda, p, q, \beta)$ and recall $M=\hat{M}(\beta)$. 

\begin{figure}[htbp]
\centering
\begin{subfigure}[b]{0.45\textwidth}
\includegraphics[width=\textwidth]{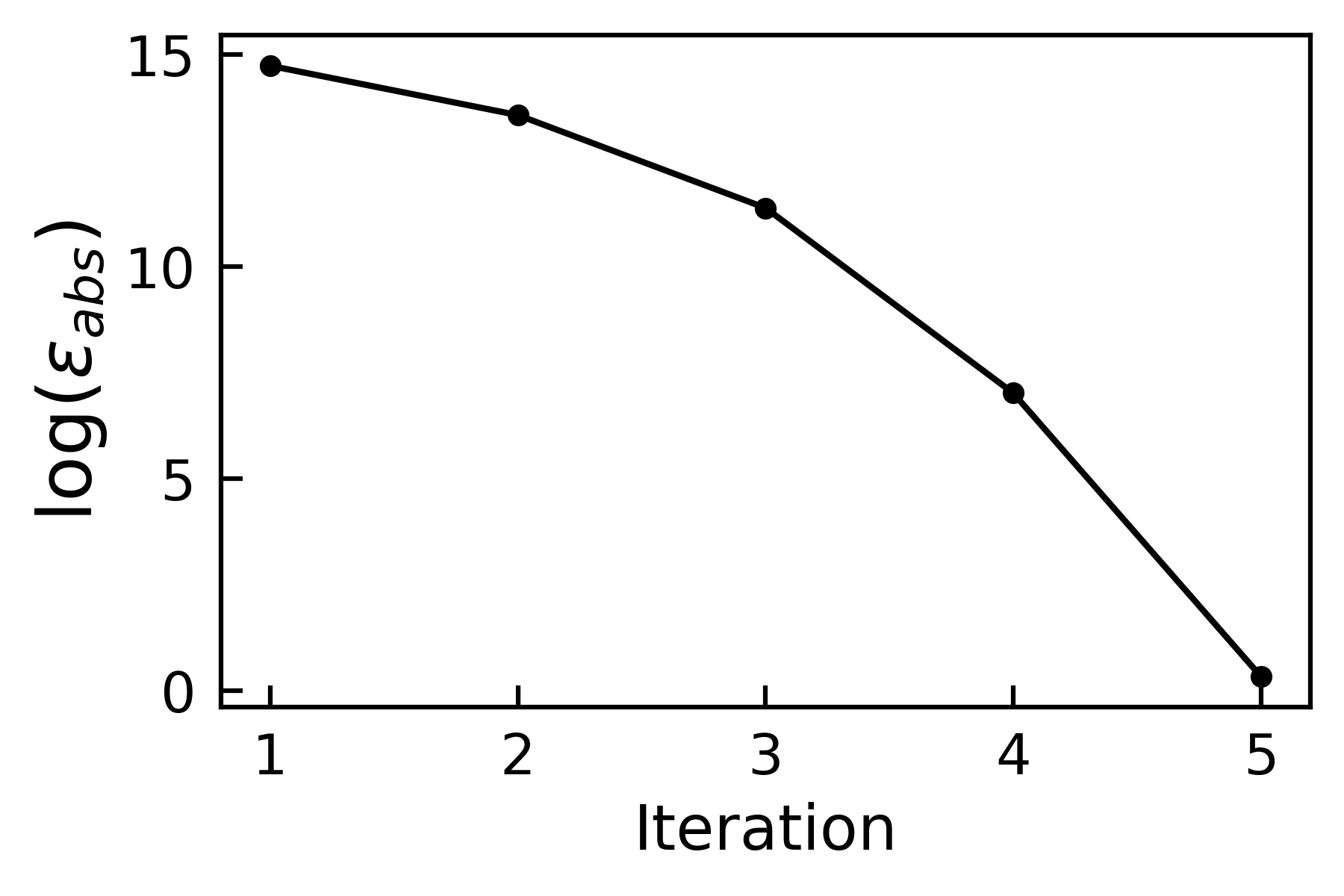}%press_prop_0s.png}
\caption{Absolute error}
\end{subfigure} %end subfigure
\hfill
\begin{subfigure}[b]{0.45\textwidth}
\includegraphics[width=\textwidth]{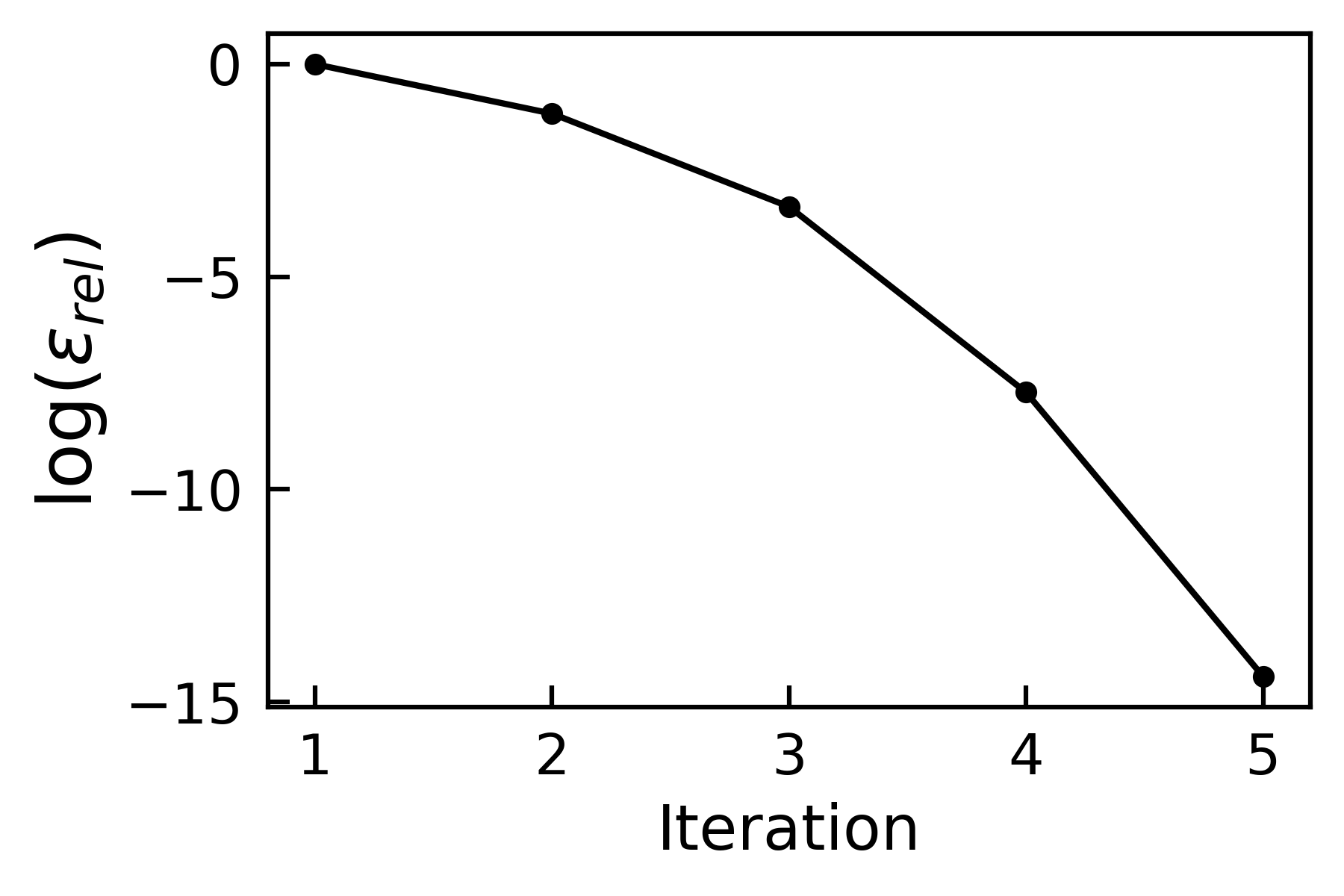}%mises_prop_0s.png}
\caption{Relative error}
\end{subfigure} %end subfigure
\caption{\blue{Example quadratic convergence of Newton-Raphson iteration for absolute error ($R$) and relative error ($R/R_0$, where $R_0$ is the residual based upon the trial state). The algorithm was tested in a large variety of yielding stress states and at various points on the yield surface, for which quadratic convergence of the iteration was always observed. This included paths through the Yp-MCC surface transition point, the preconsolidation intercept, and many others, as well as in large-mesh simulations where stress states and paths are variable. We could not get the local material Newton-Raphson iteration to fail in any of these tests.}}
\label{fig:NRconvergence}
\end{figure}

\blue{The Newton-Raphson iteration converges quadratically (Fig.\ref{fig:NRconvergence}), tested for a wide variety of load paths and stress states, as well as in large mesh simulations where paths and states vary from integration point to point. The details of the equations for the sake of reproducibility are given in the following.} The Jacobian is found $J := \nabla_x R$. We note the following relevant sub-derivatives, beginning with the fourth element of the residual vector,
\beq \label{eq:jacobian_start}
\dd{R_4}{\beta} = 1 -H\dd{\Delta z}{\beta} - \beta_{max} \omega \exp{\omega z_{(n)}}\exp{\omega \Delta z}\dd{\Delta z}{\beta},
\eeq
where
\beq
\dd{\Delta z}{\beta} = \dl (\frac{4q^2}{M^3}\frac{Xp_{c0}M_0}{p_c^3} - \frac{6q^2}{M^4}\left(\frac{Xp_{c0}M_0}{p_c^2}   \right)^2.
\eeq

\beq
\dd{R_4}{q} = -H\dd{\Delta z}{q} - \beta_{max} \omega \exp{\omega z_{(n)}} \exp{\omega \Delta z} \dd{\Delta z}{q},
\eeq
where 
\beq
\dd{\Delta z}{q} = \dl\left(\frac{4q}{M^3} \frac{Xp_{c0}M_0}{p_c^2}  \right).
\eeq

\beq
\dd{R_4}{p} = -H\dd{\Delta z}{p} - \beta_{max} \omega \exp{\omega z_{(n)}} \exp{\omega \Delta z} \dd{\Delta z}{p},
\eeq
where
\beq
\dd{\Delta z}{p} = -\dl.
\eeq

\beq
\dd{R_4}{\mu} =\dd{R_4}{\dl}\dd{\dl}{\mu}= -\frac{H}{G_{nr}}\dd{\Delta z}{\dl} - \frac{\beta_{max} \omega}{G_{nr}} \exp{\omega z_{(n)}} \exp{\omega \Delta z} \dd{\Delta z}{\dl},
\eeq
where
\beq
\dd{\Delta z}{\dl} = \frac{2q^2}{M^3} \frac{Xp_{c0}M_0}{p_c^2} - p.
\eeq

Then for the third element of the residual vector,
\beq
\dd{R_3}{\beta} = \frac{-12q\mu}{ M^3} \dd{M}{\beta},
\eeq
where
\beq
\dd{M}{\beta} = \frac{Xp_{c0}M_0}{p_c^2}.
\eeq

\beq
\dd{R_3}{q} = 1 +6\mu/M^2
\eeq

\beq
\dd{R_3}{p} = 0.
\eeq

\beq
\dd{R_3}{\mu} = 6q/M^2.
\eeq

For the second element of the residual vector,
\beq
\dd{R_2}{\beta} = \frac{K_{nr}\mu}{G_{nr}},
\eeq

\beq
\dd{R_2}{q} = 0,
\eeq

\beq
\dd{R_2}{p} = 1+2\mu K_{nr}/G_{nr},
\eeq

\beq
\dd{R_2}{\mu} = (2p-pc)K_{nr}/G_{nr} .
\eeq

And, lastly, for the first element of the residual vector, such that
\beq
\dd{R_1}{\beta} = p - \frac{2q^2}{M^3} \dd{M}{\beta} ,
\eeq

\beq
\dd{R_1}{q} = \frac{2q}{M^2}
\eeq

\beq
\dd{R_1}{p} = 2p -p_c,
\eeq

\beq \label{eq:jacobian_end}
\dd{R_1}{\mu} = 0.
\eeq

\bibliographystyle{elsarticle-harv}
\addcontentsline{toc}{section}{\refname}\bibliography{bib}

\end{document}